\documentclass{article}

\usepackage[utf8]{inputenc} 
\usepackage[T1]{fontenc}    
\usepackage{amsfonts}       
\usepackage{amsmath}
\usepackage{booktabs}       
\usepackage{cite}
\usepackage{float}
\usepackage{graphicx}
\usepackage{hyperref}       
\usepackage{lipsum}
\usepackage{microtype}      
\usepackage{multirow}
\usepackage{nicefrac}       
\usepackage{url}            

\title{High-dimensional Statistics Applications to Batch Effects in Metabolomics}

\author{Zhendong Guo\\
        Fudan University\\
        \texttt{guozhendong19@mails.ucas.ac.cn}}
\date{December 13, 2024}

\begin{document}
\maketitle
\begin{abstract}
Batch effects are inevitable in large-scale metabolomics. Prior to formal data analysis, batch effect correction (BEC) is applied to prevent from obscuring biological variations, and batch effect evaluation (BEE) is used for correction assessment. However, existing BEE algorithms neglect covariances between the variables, and existing BEC algorithms might fail to adequately correct the covariances. Therefore, we resort to recent advancements in high-dimensional statistics, and respectively propose “quality control-based simultaneous tests (QC-ST)” and “covariance correction (CoCo)”. Validated by the simulation data, QC-ST can simultaneously detect the statistical significance of QC samples' mean vectors and covariance matrices across different batches, and has a satisfactory statistical performance in empirical sizes, empirical powers, and computational speeds. Then, we apply four QC-based BEC algorithms to two large cohort datasets, and find that extreme gradient boost (XGBoost) performs best in relative standard deviation (RSD) and dispersion-ratio (D-ratio). After prepositive BEC, if QC-ST still suggests that batch effects between some two batches are significant, CoCo should be implemented. And after CoCo (if necessary), the four metrics (i.e., RSD, D-ratio, classification performance, and QC-ST) might be further improved. In summary, under the guidance of QC-ST, we can develop a matching strategy to integrate multiple BEC algorithms more rationally and flexibly, and minimize batch effects for reliable biological conclusions.
\end{abstract}

\section{Introduction}
Metabolomics is an important component of systems biology, focusing on small molecule metabolites ($<1500 \text{ Da}$). Data acquisition in metabolomics commonly employs automated analytical instruments such as liquid chromatography (LC), gas chromatography (GC), mass spectrometry (MS), and nuclear magnetic resonance (NMR) spectroscopy. However, large-scale metabolomics necessitates long-term and multi-batch experimental procedures leading to technical variations (i.e., batch effects) inevitably, while biological variations are the focus of attention \cite{ref1, ref2, ref3}.

Batch effects can be categorized into intra-batch effects (e.g., time-dependent drift in instrumental response) and inter-batch effects (e.g., inconsistent operations, reagents, and instrumental conditions), both of which will reduce statistical powers, limit repeatability and reproducibility, and obscure biological variations \cite{ref4}. Therefore, prior to formal data analysis, it is imperative to apply batch effect correction (BEC) algorithms to attain reliable biological conclusions. BEC algorithms can be categorized into three main types \cite{ref2}: data-driven methods, internal standard (IS)-based methods, and quality control (QC)-based methods. Among them, QC-based methods are the most popular in metabolomics. This is because QC samples are typically pooled from all the subject samples, which can also equilibrate the instrumental system and evaluate the experimental precision \cite{ref1, ref3, ref5}. QC-based BEC algorithms include: batch-ratio \cite{ref6, ref7, ref8}, linear regression \cite{ref6, ref8}, locally estimated scatter plot smoothing (LOESS) regression \cite{ref9}, smoothing spline regression \cite{ref10}, support vector regression (SVR) \cite{ref11, ref12, ref13}, random forest (RF) regression \cite{ref14, ref15}, technical variation elimination with ensemble learning architecture (TIGER) \cite{ref16}, etc. Inspired by them, we also use extreme gradient boost (XGBoost) regression \cite{ref17, ref18}, a prominent algorithm recently, for BEC in this paper.

To assess the performance of BEC and integrate the data across different batches, batch effect evaluation (BEE) is also essential. BEE algorithms can be categorized into qualitative methods and quantitative methods \cite{ref2, ref19}. The former include: principal component analysis (PCA), and hierarchical clustering analysis (HCA). The latter include: principal variance component analysis (PVCA) \cite{ref20}, the stick plot \cite{ref21}, guided PCA (gPCA) \cite{ref22}, and signal-to-noise ratio (SNR) \cite{ref23}. PVCA and the stick plot fail to provide clear cutoffs. SNR provides an empirical cutoff (e.g., 10), and is only feasible when the sample size of each batch is equal. In contrast, gPCA is a statistical method based on the permutation test and determines the significance of batch effects by the significance level $\alpha_{\mathrm{sig}}=0.05$. Compared to BEC algorithms, BEE algorithms are less studied and have more challenges to develop. This is because the QC sample size of each batch (denoted as $n$) is usually smaller than the variable number (denoted as $p$) in metabolomics, namely the “large $p$, small $n$” problem. Consequently, traditional multivariate statistical methods (e.g., Hotelling's $T^2$ test, and Box's $M$ test) \cite{ref24} are incompetent. Fortunately, great progress has been made in high-dimensional statistics over the past 20 years \cite{ref25}. Notable advancements include: Chen and Qin's homogeneity test of mean vectors \cite{ref26}, and Li and Chen's homogeneity test of covariance matrices \cite{ref27}.

Considering the homogeneity of all the QC samples, we assume that if without batch effects, QC samples across different batches should follow the same multivariate normal distribution. Therefore, we resort to the simultaneous test of mean vectors (denoted as $\boldsymbol{\mu}$) and covariance matrices (denoted as $\boldsymbol{\Sigma}$) \cite{ref28, ref29, ref30}, and propose a new BEE algorithm termed “QC-based simultaneous tests (QC-ST)”. Then, we apply four BEC algorithms (i.e., SVR, RF, TIGER, and XGBoost) to two large cohort datasets, and find that they might fail to adequately correct the covariance matrices. To overcome the limitations, we also propose a new BEC algorithm termed “covariance correction (CoCo)”, which employs the graphical elastic net (GELNET) theory for high-dimensional normal data. Finally, we select four metrics for correction assessment, i.e., relative standard deviation (RSD), dispersion-ratio (D-ratio), classification performance, and QC-ST. The relevant functions have been encapsulated into an R package termed “\texttt{MetBEC}”, which can be downloaded from 
\begin{center}
    \url{https://github.com/Bubble-o0O/MetBEC}.
\end{center}

\section{Methods}
\subsection{Statistical Performance Assessment}
Currently, there are nine simultaneous test methods \cite{ref28, ref29, ref30}. However, considering that the QC sample size of each batch is not large, effective methods for small sample sizes should be screened out. The null hypothesis of the simultaneous test is: 
\begin{equation*}
    H_0: \boldsymbol{\mu}_{1}=\boldsymbol{\mu}_{2}, \boldsymbol{\Sigma}_{1}=\boldsymbol{\Sigma}_{2}.
\end{equation*}
And the alternative hypothesis is divided into three types: 
\begin{equation*}
    H_1:
    \begin{cases}
        H_\mathrm{m}: \boldsymbol{\mu}_{1}\neq\boldsymbol{\mu}_{2}, \boldsymbol{\Sigma}_{1}=\boldsymbol{\Sigma}_{2}; \\ H_\mathrm{c}: \boldsymbol{\mu}_{1}=\boldsymbol{\mu}_{2}, \boldsymbol{\Sigma}_{1}\neq\boldsymbol{\Sigma}_{2}; \\ H_\mathrm{m} \cap H_\mathrm{c}: \boldsymbol{\mu}_{1}\neq\boldsymbol{\mu}_{2}, \boldsymbol{\Sigma}_{1}\neq\boldsymbol{\Sigma}_{2}.
    \end{cases}
\end{equation*}
Choose $\alpha_{\mathrm{sig}}=0.05$, the sample sizes $n_1=n_2\in\{5,10,20,40\}$, and the variable number $p\in\{50,100,250,500\}$.

\subsubsection{Empirical Sizes}
Randomly generate two groups of samples $\boldsymbol{X}_1$ and $\boldsymbol{X}_2$, which both follow the standard multivariate normal distribution $\mathrm{N}_p(\boldsymbol{0},\boldsymbol{I}_p)$, where $\boldsymbol{I}_p$ denotes the $p \times p$ identity matrix. Repeat each $(n_1,n_2,p)$ setting 5,000 times to calculate the corresponding empirical size.

\subsubsection{Empirical Powers}
We refer to Miao et al.'s \cite{ref29} and Yu et al.'s \cite{ref30} study, and also consider the data characteristics of QC samples in metabolomics (see Appendix \nameref{Appendix A}). Randomly generate two groups of samples such that $\boldsymbol{X}_1\sim\mathrm{N}_p(\boldsymbol{\mu}_1,\boldsymbol{\Sigma}_1)$, $\boldsymbol{X}_2\sim\mathrm{N}_p(\boldsymbol{\mu}_2,\boldsymbol{\Sigma}_2)$, and autoscale (standardize) the combined data of both. Repeat each $(n_1,n_2,p)$ setting 5,000 times to calculate the corresponding empirical power. For comparison, we also assess the statistical performance of gPCA with autoscaling and 1,000 times of permutations by default.

\subsection{Dataset Information}
\subsubsection{Data Pre-processing}
The modified 80\% rule \cite{ref31} is used for variable selection. \texttt{MissForest} \cite{ref32} is used for missing value imputation. Hotelling's $T^2$ statistic and DModX \cite{ref33, ref34} by PCA are used for QC samples' outlier detection, where the Tracy-Widom distribution is used to determine the principal component number \cite{ref35}. After data pre-processing, the basic information of two large cohort datasets is shown in Table \ref{Table 1}.
\begin{table}[H]
    \centering
    \caption{Dataset information}
    \begin{tabular}{cccccc}
    \toprule
        Dataset &
        \begin{tabular}[c]{@{}c@{}}
            Batch \\  
            number 
        \end{tabular} &
        \begin{tabular}[c]{@{}c@{}}
            Variable \\  
            number 
        \end{tabular} &
        \begin{tabular}[c]{@{}c@{}}
            Total \\ 
            sample size
        \end{tabular} &
        \begin{tabular}[c]{@{}c@{}}
            QC \\
            sample size
        \end{tabular} &
        \begin{tabular}[c]{@{}c@{}}Subject \\
            sample size
        \end{tabular} \\
    \midrule
        I & 4 & 268 & 1296 & 122 & 1174 \\
        II & 15 & 53 & 1164 & 162 & 1002 \\
    \bottomrule
    \end{tabular}
    \label{Table 1}
\end{table}

\subsubsection{Dataset I}
It was downloaded from Fan et al.'s \cite{ref15} study, and was based on plasma samples from the P20 study. Reversed-phase liquid chromatography (RPLC) coupled to quadrupole time-of-flight mass spectrometry (QTOF-MS) in negative electrospray ionization mode was used for untargeted lipidomics data acquisition, and a validated lipidomics assay was used to identify the metabolites. Dataset I has 4 batches, 268 metabolites, and 1296 samples, where the QC sample sizes are respectively 31, 32, 31, 28, and the subject sample sizes are respectively 303, 303, 303, 265.

\subsubsection{Dataset II}
It was downloaded from Kim et al.'s \cite{ref36} study, and was based on plasma samples from the BioHEART-CT study. Hydrophilic interaction chromatography (HILIC) coupled to AB SCIEX QTARP 5500 mass spectrometry was used for targeted metabolomics data acquisition, and pure compound was used to identify the metabolites. Dataset II has 15 batches, 53 metabolites, and 1164 samples, where the QC sample sizes are respectively 10, 11, 11, 11, 10, 11, 11, 10, 11, 11, 11, 11, 11, 11, 11, and the subject sample sizes are respectively 73, 66, 62, 70, 66, 66, 66, 65, 66, 66, 67, 66, 66, 66, 71. Among the subject samples, 390 individuals have hypertension.

\subsection{BEC}
We use SVR, RF, TIGER, and XGBoost respectively for intra-BEC, and use batch-ratio for inter-BEC.

SVR, RF, and XGBoost have the same correction principle. Specifically, establish the regression model batchwise \cite{ref12, ref13, ref15, ref16}:
\begin{equation*}
    y_i \sim f(x,\boldsymbol{y}_{(i)}\mid\boldsymbol{\theta}),i=1,2,...,p,
\end{equation*}
where $y_i$ denotes the intensity of QC samples' $i$th metabolite; $x$ denotes the QC samples' injection order; $\boldsymbol{y}_{(i)}$ denotes the several (10 by default) variables with the highest correlations to the $i$th variable; $\boldsymbol{\theta}$ denotes the optimal hyperparameters. TIGER has a more complex correction principle, which can be divided into two parts: the base model and the meta model. The former employs RF to establish the regression model similar to the above, and the latter integrates several base models with different hyperparameters weighted by the loss function (see Han et al.'s \cite{ref16} study for details). As for batch-ratio, it has several versions \cite{ref6, ref7, ref8}. Here, we recommend Wang et al.'s \cite{ref8} version (also termed “ratio-A”) due to the combination of median values and mean values.
\begin{figure}[H]
    \centering
    \includegraphics[width=1\linewidth]{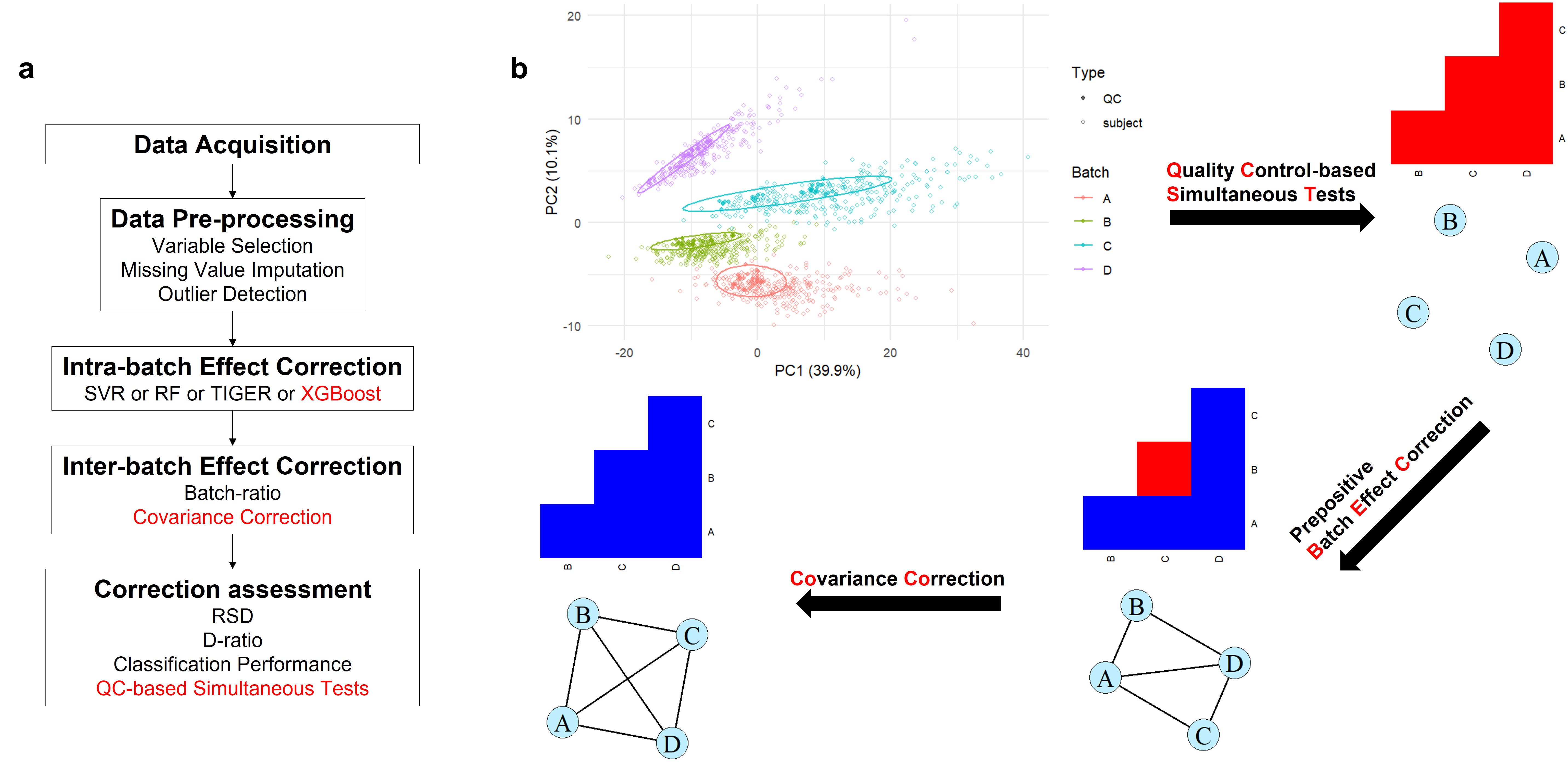}
    \caption{Overview}
    \footnotesize

    \textbf{a}: Workflow of data processing. \textbf{b}: Visualization by the PCA score plot, the heatmap, and the undirected graph.
    \label{Figure 1}
\end{figure}

\subsection{QC-ST}
Autoscale the combined QC samples' data from all the batches, and evaluate the significance of batch effects by the simultaneous test pairwise. Choose $\alpha_{\mathrm{sig}}=0.05$, and calculate the adjusted $p\text{-value}$ (i.e., $q\text{-value}$) with the false discovery rate (FDR) method \cite{ref37}. If $q\text{-value}\ge\alpha_{\mathrm{sig}}$, batch effects between the two batches are not significant. Otherwise, if $q\text{-value}<\alpha_{\mathrm{sig}}$, batch effects between the two batches are significant, and at least one parameter has statistical significance. Then, the homogeneity test of mean vectors \cite{ref26} and the homogeneity test of covariance matrices \cite{ref27} can be used to further determine which parameter does.

\subsection{CoCo}
\subsubsection{Preliminary Knowledge}
For an $n \times p$ mean-centered data $\boldsymbol{X}$, the empirical covariance matrix is given by $\mathrm{Cov}(\boldsymbol{X})=\boldsymbol{S}=\frac{1}{n-1}\boldsymbol{X}^{\mathrm{T}}\boldsymbol{X}$. When $n_1,n_2>p$, if making the empirical covariance matrices of the mean-centered data $\boldsymbol{X}_1$ and $\boldsymbol{X}_2$ after correction equal (denoted as $\tilde{\boldsymbol{\Sigma}}$), we should let $\boldsymbol{Y}_1=\boldsymbol{X}_1\boldsymbol{A}_1$, $\boldsymbol{Y}_2=\boldsymbol{X}_2\boldsymbol{A}_2$, where the transformation matrices are given by $\boldsymbol{A}_1=\boldsymbol{S}_1^{-\frac{1}{2}} \tilde{\boldsymbol{\Sigma}}^{\frac{1}{2}}$, $\boldsymbol{A}_2=\boldsymbol{S}_2^{-\frac{1}{2}} \tilde{\boldsymbol{\Sigma}}^{\frac{1}{2}}$ such that $\mathrm{Cov}(\boldsymbol{Y}_1 )=\mathrm{Cov}(\boldsymbol{Y}_2 )=\tilde{\boldsymbol{\Sigma}}$.

However, the “large $p$, small $n$” problem leads to non-invertible empirical covariance matrices, hence the above transformation matrices can't be directly obtained. To overcome the limitations, we resort to the Gaussian graphical model (GGM) theory. For the multivariate normal distribution $\mathrm{N}_p(\boldsymbol{\mu},\boldsymbol{\Sigma})$, the maximum likelihood estimator (MLE) of $\boldsymbol{\Sigma}$ is given by $\boldsymbol{S}=\frac{1}{n}\boldsymbol{X}^{\mathrm{T}}\boldsymbol{X}$; Only when $n>p$, the MLE of the precision matrix (inverse covariance matrix) $\boldsymbol{\Theta}=\boldsymbol{\Sigma}^{-1}$ exists, and is given by
\begin{equation*}
    \hat{\boldsymbol{\Theta}} = \arg\min_{\boldsymbol{\Theta}}\{-\mathrm{log}|\boldsymbol{\Theta}| + \mathrm{tr}(\boldsymbol{S}\boldsymbol{\Theta})\}.
\end{equation*}
To estimate the precision matrix of high-dimensional normal data, Friedman et al. \cite{ref38} proposed the graphical lasso by adding the $L_1$-penalty term into the MLE of $\boldsymbol{\Theta}$. Subsequently, Wieringen et al. \cite{ref39} and Kuismin et al. \cite{ref40} proposed the graphical ridge by adding the $L_2$-penalty term. Recently, Kovacs et al. \cite{ref41} proposed the graphical elastic net (GELNET) by combining both, such that the precision matrix estimator is given by
\begin{equation*}
    \hat{\boldsymbol{\Theta}} = \arg\min_{\boldsymbol{\Theta}}\{-\mathrm{log}|\boldsymbol{\Theta}| + \mathrm{tr}(\boldsymbol{S}\boldsymbol{\Theta}) + \lambda(\alpha\|\boldsymbol{\Theta}-\boldsymbol{T}\|_{1} + \frac{1-\alpha}{2}\|\boldsymbol{\Theta}-\boldsymbol{T}\|^2_{2})\},
\end{equation*}
where the hyperparameters $\alpha\in[0,1]$, $\lambda\in(0,+\infty)$; the target matrix $\boldsymbol{T}$ is a positive semi-definite matrix; $\|.\|_1$ denotes the matrix $L_1$-norm; $\|.\|_2$ denotes the matrix $L_2$-norm (Frobenius norm). Therefore, we can obtain the invertible covariance matrix estimator $\hat{\boldsymbol{\Sigma}}=\hat{\boldsymbol{\Theta}}^{-1}$ of high-dimensional normal data by GELNET.

\subsubsection{Algorithm}
\begin{enumerate}
    \item{
    Suppose that the batch number is denoted as $B$, and the QC sample sizes are batchwise denoted as $n_j,j=1,2,...,B$. Firstly, we obtain the QC samples' covariance matrices $\hat{\boldsymbol{\Sigma}}_j=\hat{\boldsymbol{\Theta}}_j^{-1}$ by GELNET. Then, let the transformation matrices
        \begin{equation*}
            \boldsymbol{A}_j=\hat{\boldsymbol{\Sigma}}_j^{-\frac{1}{2}} \tilde{\boldsymbol{\Sigma}}^{\frac{1}{2}}=\hat{\boldsymbol{\Theta}}_j^{\frac{1}{2}} \tilde{\boldsymbol{\Sigma}}^{\frac{1}{2}},
        \end{equation*}
    where the normalized covariance matrix is given by
        \begin{equation*}
            \tilde{{\boldsymbol{\Sigma}}}=\frac{\sum_{j=1}^{B}n_j\hat{\boldsymbol{\Sigma}}_j}{\sum_{j=1}^{B}n_j}.
        \end{equation*}
    }
    \item{
    Let the transformation matrices be applied to subject samples of the corresponding batch, namely
    \begin{equation*}
        \boldsymbol{Y}_j^{\mathrm{(ss)}}=\boldsymbol{X}_j^{\mathrm{(ss)}}\boldsymbol{A}_j,
    \end{equation*}
    where $\boldsymbol{X}_j^{\mathrm{(ss)}}$ denotes the subject samples' mean-centered data of the $j$th batch. 
    }
    \item{
    Translate $\boldsymbol{Y}_j^{\mathrm{(ss)}}$ to make the mean vector equal to that of $\boldsymbol{X}_j^{\mathrm{(ss)}}$ before mean-centering.
    }
    \item{
    Hyperparameter optimization: \\
    Suppose $\boldsymbol{\alpha}=(\alpha_1,\alpha_2,...,\alpha_B )^{\mathrm{T}}$, $\boldsymbol{\lambda}=(\lambda_1,\lambda_2,...,\lambda_B )^{\mathrm{T}}$. According to the relevant literature \cite{ref41, ref42, ref43}, let $\alpha_j\in[0,1]$, $\lambda_j\in(0,10)$, $\boldsymbol{T}_j=\boldsymbol{I}_{p}$, $j=1,2,...,B$. Subsequently, use 500 times of random search (i.e., obtain 500 $(\boldsymbol{\alpha},\boldsymbol{\lambda})$ settings randomly) by default, and select the $(\boldsymbol{\alpha},\boldsymbol{\lambda})$ setting which meets the following conditions successively:
    \begin{itemize}
        \item{
        There must be no statistical significance between the QC samples' covariance matrices of any two batches. Suppose $u$ $(\boldsymbol{\alpha},\boldsymbol{\lambda})$ settings which meet this condition;
        } 
        \item{
        To prevent overcorrection for the subject samples, firstly, calculate their variance fold changes of each variable before and after CoCo batchwise (i.e., obtain a $B \times p$ matrix of the variance fold changes denoted as $\boldsymbol{V}$). Then, calculate the mean value of the matrix, namely $\mathrm{mean}(\boldsymbol{V})$. Finally, repeat $u$ times from the above settings, and select the $(\boldsymbol{\alpha},\boldsymbol{\lambda})$ setting with the minimum $\mathrm{mean}(\boldsymbol{V})$.
        } 
    \end{itemize}
    }
\end{enumerate}

\subsection{Correction Assessment}
Four metrics are selected, i.e., RSD, D-ratio, classification performance, and QC-ST. Among them, RSD and QC-ST only examine QC samples, which might lead to overoptimistic and incomplete results. D-ratio proposed by Broadhurst et al. \cite{ref5} examines both QC samples and subject samples, whose acceptance criterion is 50\%. Classification performance only examines subject samples, whose details are as follows \cite{ref15, ref23, ref44}: Firstly, we combine the Mann-Whitney U test with FDR adjustment, and the variable importance projection (VIP) \cite{ref45} of partial least squares-discriminant analysis (PLS-DA). Variables with $q\text{-value}<0.05$ and $\text{VIP-value}>1$ are selected as biomarkers \cite{ref34}. Then, we use the gradient boosting machine (GBM) classifier \cite{ref17} with 99 times of down-sampling to calculate the area under curve (AUC) of receiver operating characteristic (ROC) and the Matthews correlation coefficient (MCC).

The formulae of RSD and D-ratio are respectively:
\begin{equation*}
    \mathrm{RSD}_i=\frac{\sqrt{\mathrm{Var}(y^{\mathrm{(QC)}}_i)}}{\mathrm{mean}(y^{\mathrm{(QC)}}_i)},
\end{equation*}
\begin{equation*}
    \mathrm{D\text{-}ratio}_i=\sqrt{\frac{\mathrm{Var}(y^{\mathrm{(QC)}}_i)}{\mathrm{Var}(y^{\mathrm{(QC)}}_i) + \mathrm{Var}(y^{\mathrm{(ss)}}_i)}},
\end{equation*}
where $i=1,2,...,p$; $y^{\mathrm{(QC)}}_i$ denotes the intensity of QC samples' $i$th metabolite; $y^{\mathrm{(ss)}}_i$ denotes the intensity of subject samples' $i$th metabolite. 

\section{Results and Discussion}
\subsection{Statistical Performance Assessment}
\subsubsection{Empirical Sizes}
Among nine simultaneous test methods, three methods (respectively termed “HN” \cite{ref28}, “Yu-Fisher” \cite{ref30}, and “Yu-Cauchy” \cite{ref30}) outperform the remaining six methods (see Appendix \nameref{Appendix B}). Additionally, gPCA also performs well. Table \ref{Table 2} shows that their empirical sizes are in $0.04\sim0.08$, and close to the nominal significance level $\alpha_{\mathrm{sig}}=0.05$.

\begin{table}[H]
    \centering
    \caption{Empirical sizes for the null hypothesis $H_0$}
    \begin{tabular}{cccccc}
    \toprule
    \multirow{2}{*}{Method} & \multirow{2}{*}{$n_1,n_2$} & \multicolumn{4}{c}{$p$} \\
                            \cmidrule{3-6}
                            &                        & 50     & 100    & 250    & 500    \\
    \midrule
    HN                      & 5, 5                   & 0.0726 & 0.0722 & 0.0682 & 0.0636 \\
                            & 10, 10                 & 0.0664 & 0.0622 & 0.0636 & 0.0536 \\
                            & 20, 20                 & 0.0558 & 0.0570 & 0.0574 & 0.0526 \\
                            & 40, 40                 & 0.0532 & 0.0586 & 0.0520 & 0.0514 \\
    Yu-Cauchy               & 5, 5                   & 0.0562 & 0.0536 & 0.0556 & 0.0498 \\
                            & 10, 10                 & 0.0476 & 0.0486 & 0.0528 & 0.0522 \\
                            & 20, 20                 & 0.0464 & 0.0490 & 0.0478 & 0.0458 \\
                            & 40, 40                 & 0.0476 & 0.0504 & 0.0500 & 0.0558 \\
    Yu-Fisher               & 5, 5                   & 0.0524 & 0.0512 & 0.0474 & 0.0428 \\
                            & 10, 10                 & 0.0530 & 0.0474 & 0.0462 & 0.0428 \\
                            & 20, 20                 & 0.0560 & 0.0510 & 0.0508 & 0.0470 \\
                            & 40, 40                 & 0.0572 & 0.0586 & 0.0528 & 0.0506 \\
    gPCA                    & 5, 5                   & 0.0562 & 0.0536 & 0.0556 & 0.0498 \\
                            & 10, 10                 & 0.0476 & 0.0486 & 0.0528 & 0.0522 \\
                            & 20, 20                 & 0.0464 & 0.0490 & 0.0478 & 0.0458 \\
                            & 40, 40                 & 0.0476 & 0.0504 & 0.0500 & 0.0558 \\
    \bottomrule
    \end{tabular}
    \label{Table 2}
\end{table}

\subsubsection{Empirical Powers}
Table \ref{Table 3} shows the empirical powers of HN, Yu-Fisher Yu-Cauchy, and gPCA for the alternative hypothesis $H_\mathrm{m} \cap H_\mathrm{c}$. Their empirical powers for the alternative hypothesis $H_\mathrm{m}$ and $H_\mathrm{c}$ are respectively shown in Table \ref{Table 6} and Table \ref{Table 7}.

\begin{table}[H]
    \centering
    \caption{Empirical powers for the alternative hypothesis $H_\mathrm{m} \cap H_\mathrm{c}$}
    \begin{tabular}{cccccc}
    \toprule
    \multirow{2}{*}{Method} & \multirow{2}{*}{$n_1,n_2$} & \multicolumn{4}{c}{$p$} \\
                            \cmidrule{3-6}
                            &                        & 50     & 100    & 250    & 500    \\
    \midrule
HN        & 5, 5   & 35.20\%  & 45.12\%  & 53.76\%  & 61.56\%  \\
          & 10, 10 & 93.14\%  & 97.40\%  & 99.26\%  & 99.82\%  \\
          & 20, 20 & 100.00\% & 100.00\% & 100.00\% & 100.00\% \\
          & 40, 40 & 100.00\% & 100.00\% & 100.00\% & 100.00\% \\
Yu-Cauchy & 5, 5   & 32.72\%  & 38.98\%  & 41.74\%  & 47.92\%  \\
          & 10, 10 & 93.38\%  & 97.60\%  & 99.18\%  & 99.68\%  \\
          & 20, 20 & 100.00\% & 100.00\% & 100.00\% & 100.00\% \\
          & 40, 40 & 100.00\% & 100.00\% & 100.00\% & 100.00\% \\
Yu-Fisher & 5, 5   & 31.76\%  & 40.86\%  & 47.20\%  & 55.62\%  \\
          & 10, 10 & 93.86\%  & 97.80\%  & 99.38\%  & 99.78\%  \\
          & 20, 20 & 100.00\% & 100.00\% & 100.00\% & 100.00\% \\
          & 40, 40 & 100.00\% & 100.00\% & 100.00\% & 100.00\% \\
gPCA      & 5, 5   & 9.80\%   & 14.82\%  & 21.18\%  & 27.92\%  \\
          & 10, 10 & 7.04\%   & 13.72\%  & 29.08\%  & 49.82\%  \\
          & 20, 20 & 3.04\%   & 7.80\%   & 38.50\%  & 81.32\%  \\
          & 40, 40 & 0.64\%   & 2.78\%   & 43.88\%  & 96.18\%  \\
    \bottomrule
    \end{tabular}
    \label{Table 3}
\end{table}

\begin{figure}
    \centering
    \includegraphics[width=1\linewidth]{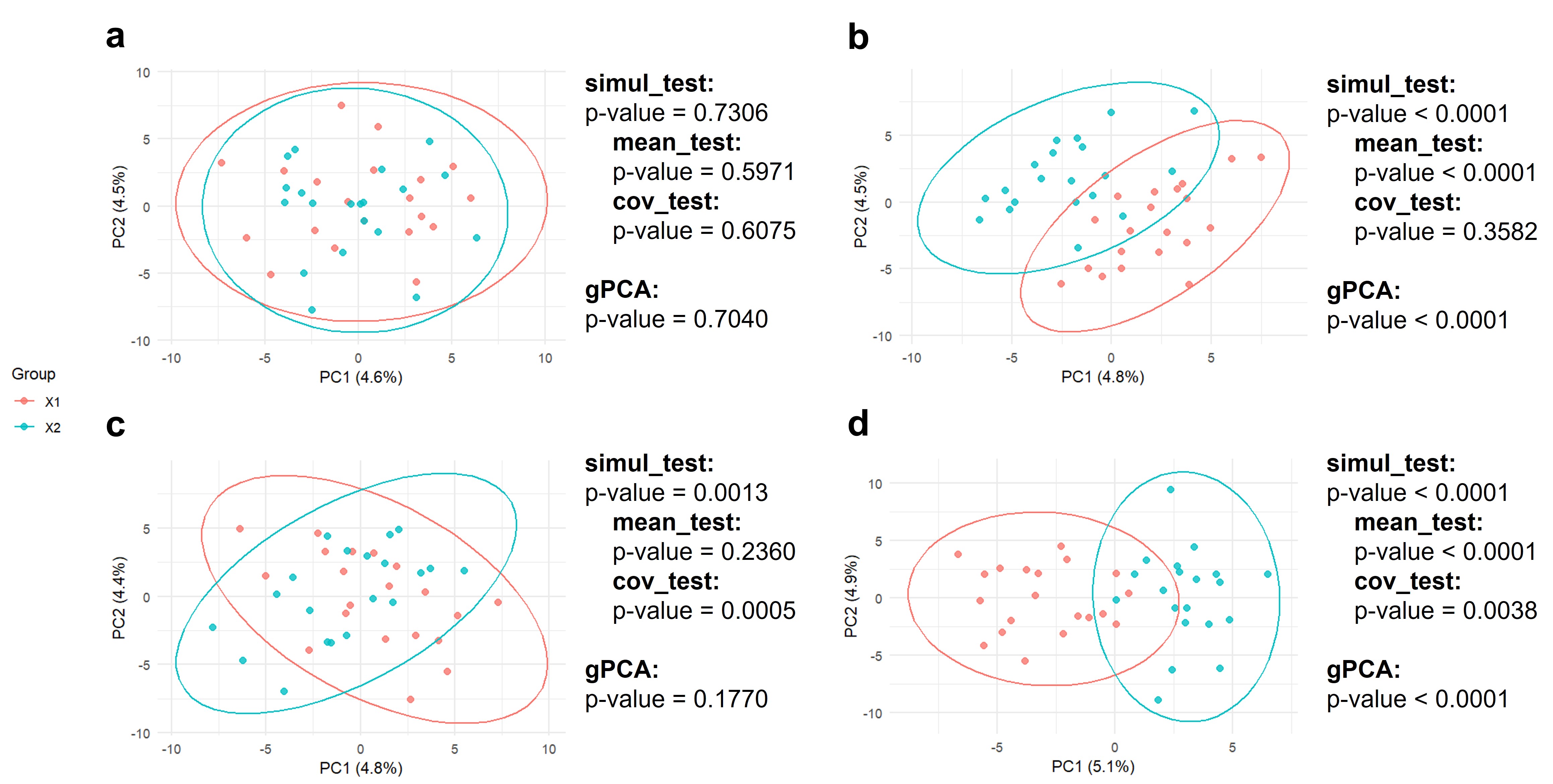}
    \caption{Statistical performance assessment}
    \footnotesize

    Here, we take $(n_1,n_2,p)=(20,20,250)$ with the Yu-Fisher method as an example. \textbf{a}: The simultaneous test and gPCA both correctly accept the null hypothesis $H_0$. \textbf{b}: The simultaneous test and gPCA both correctly accept the alternative hypothesis $H_\mathrm{m}$. \textbf{c}: The simultaneous test correctly accepts the alternative hypothesis $H_\mathrm{c}$, but gPCA makes a Type II error. \textbf{d}: The simultaneous test and gPCA both correctly accept the alternative hypothesis $H_\mathrm{m} \cap H_\mathrm{c}$.
    \label{Figure 2}
\end{figure}

Validated by the simulation data, we find that: (1) gPCA has a slower computational speed (Table \ref{Table 9}) due to employing the permutation test, and fails to detect the statistical significance of covariance matrices such that the empirical powers are lower than those of the simultaneous test methods (Figure \ref{Figure 2}c, Table \ref{Table 3} and Table \ref{Table 7}); (2) When $(n_1,n_2)$ increase from $(5,5)$ to $(10,10)$, the empirical powers of the three simultaneous test methods are obviously improved. Further, when $n_1,n_2\ge10$, their empirical powers are all larger than 80\% \cite{ref34}. Consequently, we recommend that the QC sample size of each batch should not be less than 10 to ensure high statistical powers. In practical applications, when $\mathrm{mean}\{n_1,n_2\}\ge10$, we use the Yu-Fisher method by default, otherwise we use the HN method by default.

\subsection{Correction Assessment}
\begin{table}[H]
    \centering
    \caption{The cumulative frequency (CF) of RSD and D-ratio}
    \begin{tabular}{ccccccc}
    \toprule
    \multirow{2}{*}{Dataset} & \multirow{2}{*}{Method} & \multicolumn{3}{c}{CF of RSD (\%)} &  & CF of D-ratio (\%) \\ 
                        \cmidrule{3-5} \cmidrule{7-7} 
                        &                    & $<15\%$ & $<20\%$ & $<30\%$ &  & $<50\%$ \\
    \midrule
I  & raw            & 0.75  & 3.36  & 63.06 &  & 52.99 \\
   & SVR            & 97.01 & 98.88 & 100   &  & 100   \\
   & TIGER          & 98.51 & 100   & 100   &  & 100   \\
   & RF             & 98.51 & 100   & 100   &  & 100   \\
   & RF + CoCo      & 98.51 & 100   & 100   &  & 100   \\
   & XGBoost        & 100   & 100   & 100   &  & 100   \\
II & raw            & 0     & 0     & 1.89  &  & 15.09 \\
   & SVR            & 22.64 & 35.85 & 88.68 &  & 73.58 \\
   & TIGER          & 32.08 & 43.40 & 98.11 &  & 73.58 \\
   & RF             & 33.96 & 45.28 & 98.11 &  & 73.58 \\
   & XGBoost        & 33.96 & 54.72 & 94.34 &  & 94.34 \\
   & XGBoost + CoCo & 45.28 & 90.57 & 96.23 &  & 100   \\      
    \bottomrule
    \end{tabular}
    \label{Table 4}
\end{table}

\subsubsection{Dataset I}
\begin{figure}
    \centering
    \includegraphics[width=1\linewidth]{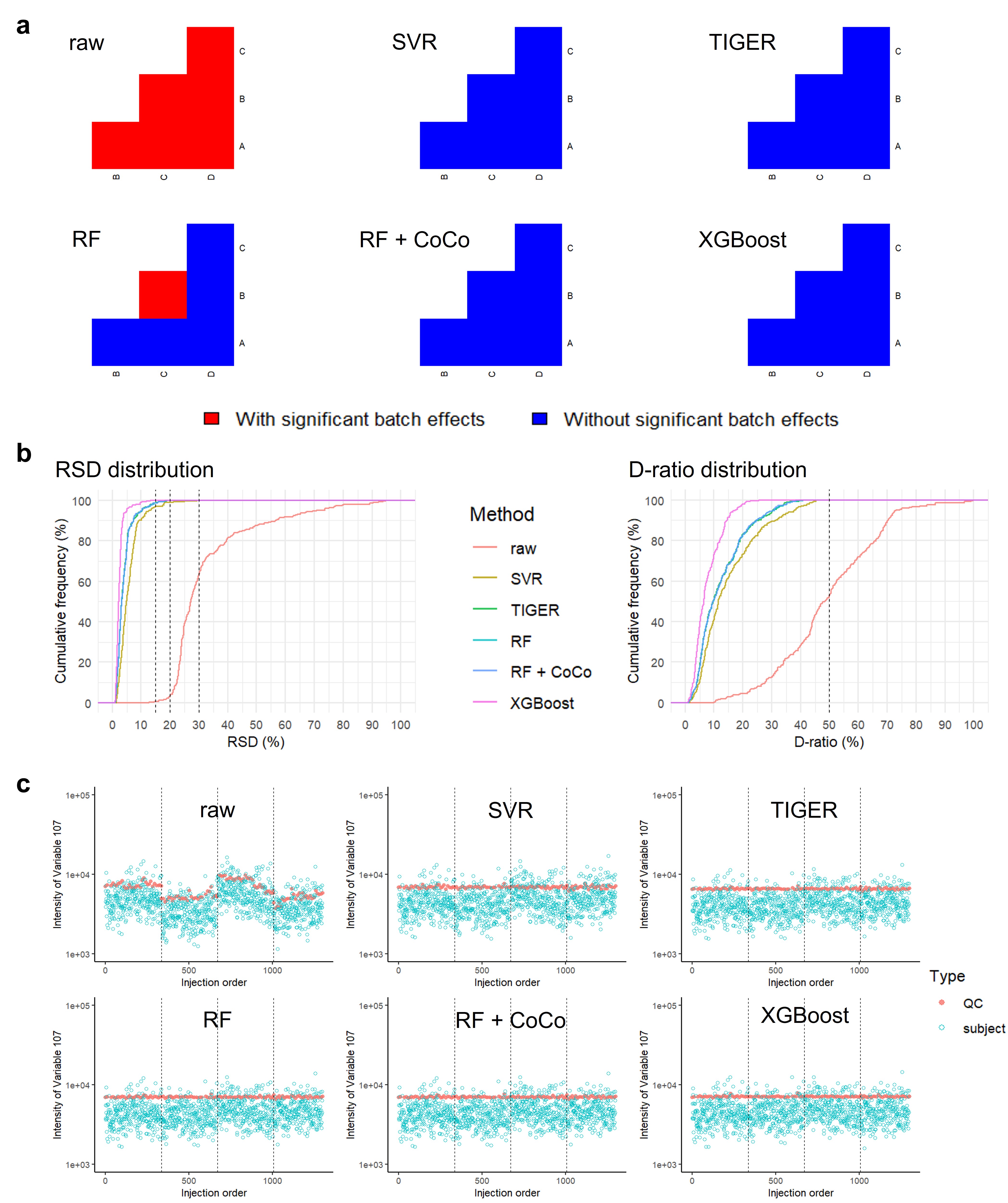}
    \caption{Correction assessment of Dataset I}
    \footnotesize

    \textbf{a}: The heatmap of QC-ST, where the capital letters denote the batch names; \textbf{b}: The distribution of RSD and D-ratio; \textbf{c}: The scatter plot of intensity, where the metabolite, plasmenyl-PC (34:2), is taken as an example here.
    \label{Figure 3}
\end{figure}

The results of QC-ST (Figure \ref{Figure 3}a) indicate that there are significant batch effects between any two batches in the raw data. After RF correction, there are still statistical significance between the covariance matrices of Batch B and Batch C, hence CoCo is implemented. As for the other three BEC algorithms, there are no significant batch effects. The results of RSD and D-ratio (Figure \ref{Figure 3}b and Table \ref{Table 4}) indicate that the distribution curve of XGBoost is located at the top left; TIGER, RF, and RF + CoCo have similar performance because their distribution curves almost overlap. Combining the above three metrics, XGBoost performs best in Dataset I.

\subsubsection{Dataset II}
\begin{figure}
    \centering
    \includegraphics[width=1\linewidth]{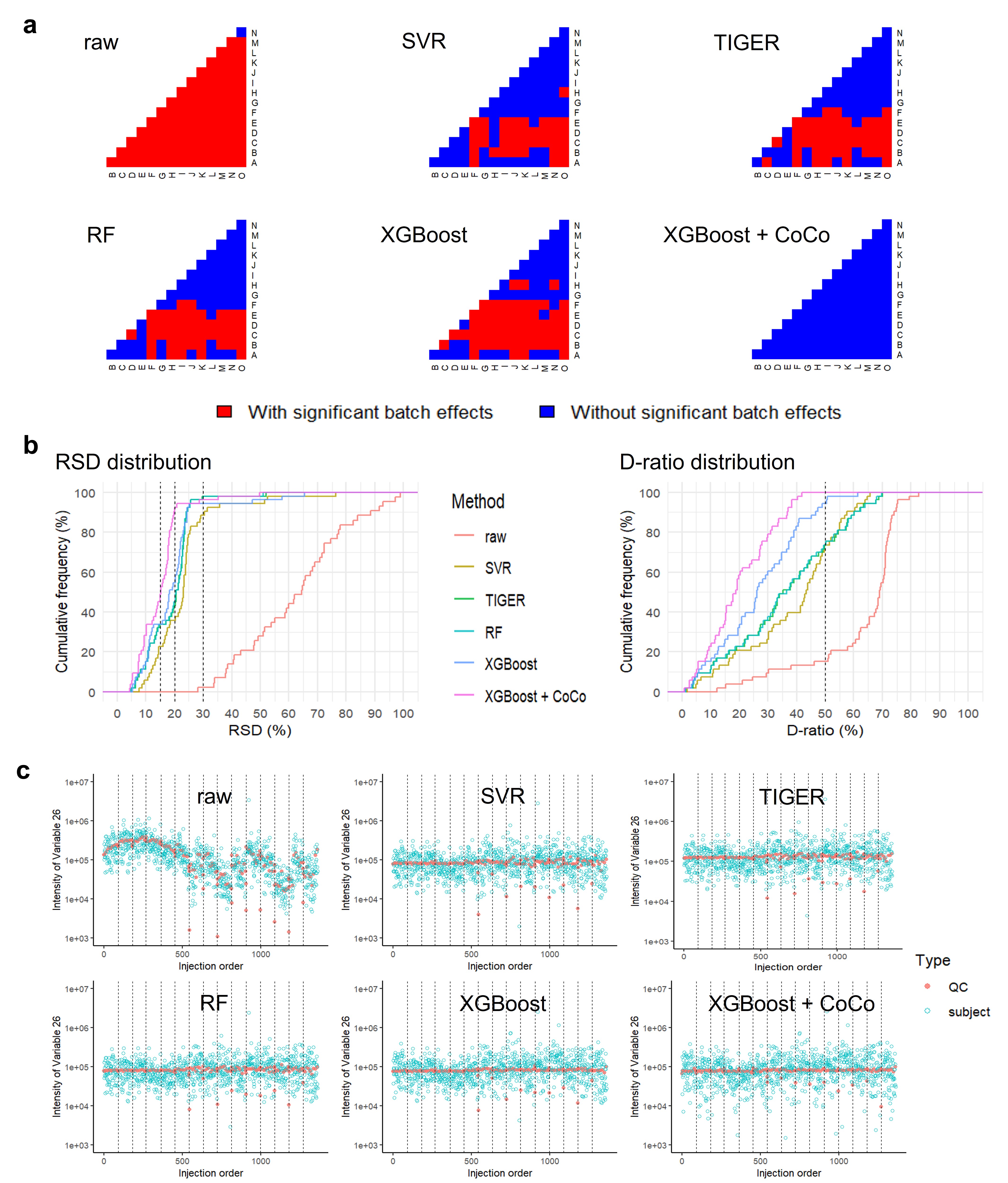}
    \caption{Correction assessment of Dataset II}
    \footnotesize

    \textbf{a}: The heatmap of QC-ST, where the capital letters denote the batch names; \textbf{b}: The distribution of RSD and D-ratio; \textbf{c}: The scatter plot of intensity, where the metabolite, dimethylguanidino valeric acid (DMGV), is taken as an example here.
    \label{Figure 4}
\end{figure}

\begin{figure}
    \centering
    \includegraphics[width=1\linewidth]{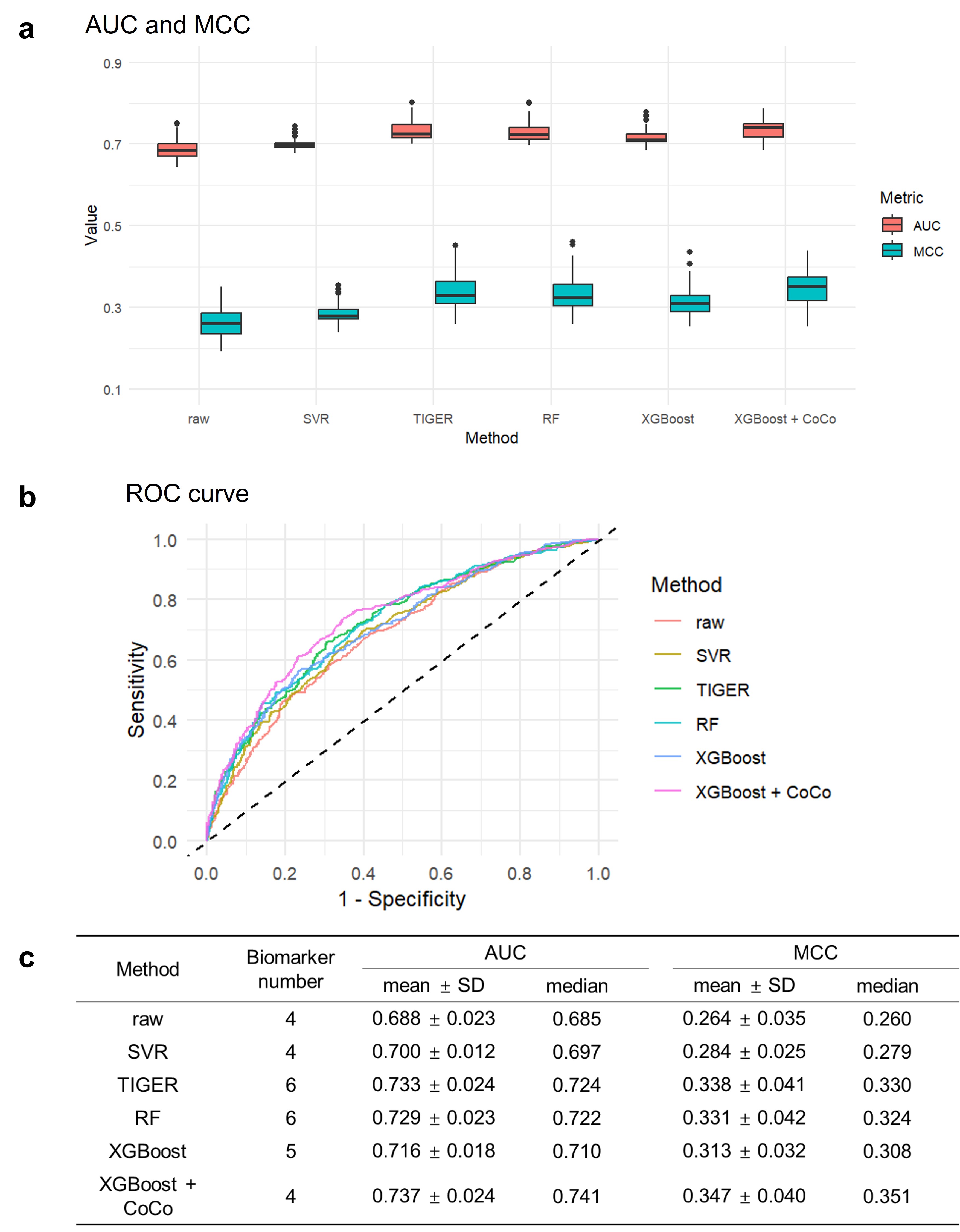}
    \caption{Classification performance of Dataset II}
    \footnotesize

    \textbf{a}: The box plot of AUC and MCC; \textbf{b}: Select the median AUC to plot the ROC curve; \textbf{c}: The descriptive statistics table of AUC and MCC with 99 times of down-sampling, where SD denotes standard deviation.
    \label{Figure 5}
\end{figure}

The results of QC-ST (Figure \ref{Figure 4}a) indicate that after correction of the four algorithms, there are still statistical significance among the covariance matrices of some batches, hence CoCo is implemented. However, only XGBoost + CoCo attains no significant batch effects, and the other three BEC algorithms fail again even after CoCo with 50,000 times of random search. The results of RSD and D-ratio (Figure \ref{Figure 4}b and Table \ref{Table 4}) indicate that the distribution curve of XGBoost + CoCo is located at the top left. The results of classification performance (Figure \ref{Figure 5}) indicate that XGBoost + CoCo has the highest AUC and MCC in terms of both mean values and median values. Combining the above four metrics, XGBoost + CoCo performs best in Dataset II.

\subsection{Discussion}
Actually, batch-ratio can ensure the consistency among QC samples' mean vectors across different batches after correction. Therefore, if QC-ST suggests that batch effects between some two batches are significant after batch-ratio correction, the covariance matrices must have statistical significance. However, there are no appropriate algorithms to correct the covariance matrices currently. Although Fan et al. \cite{ref15} have proposed that correlations between the metabolites should be considered when establishing the regression model (also termed “systematic error removal”), we find that: Even so, the statistical significance of covariance matrices might not be fully eliminated. 

CoCo employs GELNET, which can obtain the invertible covariance matrix estimator of high-dimensional normal data. Actually, if QC-ST before CoCo has indicated no significant batch effects, CoCo is no longer necessary. This is because redundant CoCo will be time-consuming, considering that the algorithm complexity is $O(Bp^3)$. Further, hyperparameter optimization of CoCo will obtain larger $\lambda_j,j=1,2,...,B,$ such that $\hat{\boldsymbol{\Theta}}_j\approx\boldsymbol{T}_j=\boldsymbol{I}_p$ and $\boldsymbol{A}_j\approx\boldsymbol{I}_p$, which means that the data before and after CoCo will keep almost unchanged. As for selection of the target matrix $\boldsymbol{T}$, several references \cite{ref40, ref41, ref42, ref43} have indicated that $\boldsymbol{T}=\boldsymbol{I}_p$ performs well, whose another advantage in CoCo is also reflected from the above analysis.

According to the results of the two datasets, compared to the raw data, RSD, D-ratio and QC-ST can be obviously improved by the four prepositive BEC algorithms, where XGBoost performs eminently. RF and TIGER perform similarly in some cases, probably because the base model of TIGER exactly employs RF. However, when the batch number is large (e.g., the batch number of Dataset II is 15), some prepositive BEC algorithms might fail again in QC-ST even after CoCo. If so, we provide several alternatives: (1) another prepositive BEC algorithm should be used (such as XGBoost); (2) some hyperparameters of the prepositive BEC algorithm should be tuned.

\section{Conclusions}
In this paper, we resort to recent advancements in high-dimensional statistics, and respectively propose a new BEE algorithm (i.e., QC-ST) and a new BEC algorithm (i.e., CoCo) based on QC samples to solve the batch effect problems in metabolomics. For BEE, existing algorithms neglect covariances between the variables. In contrast, QC-ST employs the simultaneous test of high-dimensional mean vectors and covariance matrices, and can be applied to the QC samples across different batches. For intra-BEC, there is only one reference using XGBoost in metabolomics \cite{ref16} currently. Compared to other algorithms, XGBoost has more hyperparameters when establishing the regression model, such that it is more sophisticated and powerful to handle different scales of datasets. For inter-BEC, although CoCo is time-consuming, the four metrics (RSD, D-ratio, classification performance, and QC-ST) might be further improved after CoCo. Consequently, we recommend to prioritize XGBoost for BEC, and then use CoCo if necessary.

In summary, QC-ST is competent for batch effect evaluation and correction assessment in metabolomics. Under its guidance, we can develop a matching strategy to integrate multiple BEC algorithms more rationally and flexibly (including: selection of the regression model, hyperparameter optimization, whether to use CoCo, etc.), and minimize batch effects for reliable biological conclusions.

\section*{Appendix}
\subsection*{A: The Design Scheme of Empirical Powers}
\label{Appendix A}
For mean vectors:
\begin{enumerate}
    \item{
    Let $\mu_i,i=1,2,...,p,$ all follow the uniform distribution $\mathrm{U}(0,1)$, and sort them by descending such that $\mu_1>\mu_2>...>\mu_p$;
    } 
    \item{
    Suppose $0\le\mathrm{pct}\le1$, $\eta\ge0$, and let $I=\max\{1,\lfloor\mathrm{pct} \cdot p\rfloor\}$, $\delta=\sqrt{\eta p^{-\frac{1}{2}}}$, $\mu_i'=
    \begin{cases}
        \mu_i+\delta, i=1,2,...,I \\
        \mu_i, i=I+1,I+2,...,p
    \end{cases}$;
    } 
    \item{
    Obtain the mean vectors: $\boldsymbol{\mu}_1=(\mu_1, \mu_2, ..., \mu_p )^\mathrm{T}$, $\boldsymbol{\mu}_2=(\mu_1', \mu_2', ..., \mu_p' )^\mathrm{T}$.
    } 
\end{enumerate}
For covariance matrices:
\begin{enumerate}
    \item{
    Let $\sigma_i \sim \mathrm{U}(0,0.3\mu_i)$ such that $\mathrm{RSD}_i=\frac{\sigma_i}{\mu_i}<30\%$ \cite{ref1, ref2, ref3, ref5};
    } 
    \item{
    Let the diagonal matrix $\boldsymbol{D}=\mathrm{diag}(\sigma_1, \sigma_2, ..., \sigma_p)$, the correlation matrices $\boldsymbol{R}_j=
    \begin{cases}
        (\rho_j^{|k-l|})_{1 \le k,l \le p}, 0<\rho_j\le1 \\
        \boldsymbol{I_p}, \rho_j=0
    \end{cases}, j=1,2$;
    } 
    \item{
    Obtain the covariance matrices: $\boldsymbol{\Sigma}_1=\boldsymbol{D}\boldsymbol{R}_1\boldsymbol{D}$, $\boldsymbol{\Sigma}_2=\boldsymbol{D}\boldsymbol{R}_2\boldsymbol{D}$.
    } 
\end{enumerate}

For the alternative hypothesis $H_\mathrm{m}$, choose $\mathrm{pct}=5\%$, $\eta=0.3$, $\rho_1=\rho_2=0$; For the alternative hypothesis $H_\mathrm{c}$, choose $\mathrm{pct}=\eta=0$, $\rho_1=0.3$, $\rho_2=-0.3$; For the alternative hypothesis $H_\mathrm{m} \cap H_\mathrm{c}$, choose $\mathrm{pct}=5\%$, $\eta=0.3$, $\rho_1=0.3$, $\rho_2=-0.3$.

\subsection*{B: More Results}
\label{Appendix B}
The remaining six methods are respectively termed “MX” \cite{ref29}, “MX-bootstrap” \cite{ref29}, “Yu-\texttt{chisq}” \cite{ref30}, “Yu-\texttt{pe.cauchy}” \cite{ref30}, “Yu-\texttt{pe.chisq}” \cite{ref30}, and “Yu-\texttt{pe.fisher}” \cite{ref30}. As shown in Table \ref{Table 5}, the empirical sizes of the six simultaneous test methods are obviously far from the nominal significance level $\alpha_{\mathrm{sig}}=0.05$, especially when $n$ is small. Therefore, we eliminate these methods.

Additionally, the principles of the homogeneity test of mean vectors and the homogeneity test of covariance matrices corresponding to Yu-Cauchy and Yu-Fisher are the same, but slightly different from those corresponding to HN. The homogeneity tests are respectively termed “Yu-\texttt{mean\_test}”, “Yu-\texttt{cov\_test}”, “HN-\texttt{mean\_test}”, and “HN-\texttt{cov\_test}”.
\begin{table}[H]
    \centering
    \caption{Empirical sizes for the null hypothesis $H_0$}
    \begin{tabular}{cccccc}
    \toprule
    \multirow{2}{*}{Method} & \multirow{2}{*}{$n_1,n_2$} & \multicolumn{4}{c}{$p$} \\
                            \cmidrule{3-6}
                            &                        & 50     & 100    & 250    & 500    \\
    \midrule
MX            & 5, 5   & 0.9810 & 0.9998 & 1.0000 & 1.0000 \\
              & 10, 10 & 0.4068 & 0.5702 & 0.7942 & 0.9458 \\
              & 20, 20 & 0.1250 & 0.1452 & 0.1916 & 0.2426 \\
              & 40, 40 & 0.0698 & 0.0680 & 0.0746 & 0.0776 \\
MX-bootstrap  & 5, 5   & 1.0000 & 1.0000 & 1.0000 & 1.0000 \\
              & 10, 10 & 0.6062 & 0.8420 & 0.9952 & 1.0000 \\
              & 20, 20 & 0.1606 & 0.2126 & 0.3336 & 0.4916 \\
              & 40, 40 & 0.0774 & 0.0814 & 0.1016 & 0.1110 \\
Yu-\texttt{chisq}      & 5, 5   & 0.1830 & 0.1770 & 0.1836 & 0.1778 \\
              & 10, 10 & 0.0936 & 0.0928 & 0.0962 & 0.0920 \\
              & 20, 20 & 0.0632 & 0.0622 & 0.0630 & 0.0630 \\
              & 40, 40 & 0.0560 & 0.0586 & 0.0602 & 0.0556 \\
Yu-\texttt{pe.cauchy}  & 5, 5   & 0.7814 & 0.8814 & 0.9618 & 0.9808 \\
              & 10, 10 & 0.3572 & 0.3968 & 0.4666 & 0.5420 \\
              & 20, 20 & 0.1554 & 0.1620 & 0.1446 & 0.1526 \\
              & 40, 40 & 0.1000 & 0.0932 & 0.0824 & 0.0750 \\
Yu-\texttt{pe.chisq}   & 5, 5   & 0.8244 & 0.9106 & 0.9778 & 0.9940 \\
              & 10, 10 & 0.3884 & 0.4330 & 0.4968 & 0.5724 \\
              & 20, 20 & 0.1646 & 0.1702 & 0.1566 & 0.1700 \\
              & 40, 40 & 0.0956 & 0.0922 & 0.0894 & 0.0792 \\
Yu-\texttt{pe.fisher}  & 5, 5   & 0.7896 & 0.8910 & 0.9710 & 0.9930 \\
              & 10, 10 & 0.3562 & 0.3964 & 0.4672 & 0.5424 \\
              & 20, 20 & 0.1552 & 0.1562 & 0.1450 & 0.1550 \\
              & 40, 40 & 0.0946 & 0.0906 & 0.0808 & 0.0724 \\
HN-\texttt{mean\_test} & 5, 5   & 0.0708 & 0.0700 & 0.0640 & 0.0582 \\
              & 10, 10 & 0.0598 & 0.0550 & 0.0578 & 0.0550 \\
              & 20, 20 & 0.0582 & 0.0616 & 0.0516 & 0.0508 \\
              & 40, 40 & 0.0608 & 0.0630 & 0.0604 & 0.0596 \\
HN-\texttt{cov\_test}  & 5, 5   & 0.0788 & 0.0856 & 0.0802 & 0.0722 \\
              & 10, 10 & 0.0742 & 0.0720 & 0.0758 & 0.0690 \\
              & 20, 20 & 0.0592 & 0.0584 & 0.0594 & 0.0578 \\
              & 40, 40 & 0.0638 & 0.0612 & 0.0528 & 0.0498 \\
Yu-\texttt{mean\_test} & 5, 5   & 0.0704 & 0.0690 & 0.0632 & 0.0574 \\
              & 10, 10 & 0.0592 & 0.0550 & 0.0578 & 0.0550 \\
              & 20, 20 & 0.0582 & 0.0616 & 0.0518 & 0.0508 \\
              & 40, 40 & 0.0608 & 0.0630 & 0.0604 & 0.0596 \\
Yu-\texttt{cov\_test}  & 5, 5   & 0.0286 & 0.0328 & 0.0286 & 0.0220 \\
              & 10, 10 & 0.0446 & 0.0430 & 0.0432 & 0.0402 \\
              & 20, 20 & 0.0488 & 0.0458 & 0.0456 & 0.0442 \\
              & 40, 40 & 0.0588 & 0.0538 & 0.0450 & 0.0444 \\
    \bottomrule
    \end{tabular}
    \label{Table 5}
\end{table}

\begin{table}[H]
    \centering
    \caption{Empirical powers for the alternative hypothesis $H_\mathrm{m}$}
    \begin{tabular}{cccccc}
    \toprule
    \multirow{2}{*}{Method} & \multirow{2}{*}{$n_1,n_2$} & \multicolumn{4}{c}{$p$} \\
                            \cmidrule{3-6}
                            &                        & 50     & 100    & 250    & 500    \\
    \midrule
HN            & 5, 5   & 17.30\% & 27.32\% & 32.30\%  & 42.72\%  \\
              & 10, 10 & 34.86\% & 55.98\% & 71.18\%  & 85.82\%  \\
              & 20, 20 & 67.42\% & 89.92\% & 96.64\%  & 99.82\%  \\
              & 40, 40 & 89.64\% & 99.14\% & 99.96\%  & 100.00\% \\
Yu-Cauchy     & 5, 5   & 14.64\% & 26.06\% & 33.34\%  & 47.66\%  \\
              & 10, 10 & 41.42\% & 66.70\% & 82.08\%  & 94.44\%  \\
              & 20, 20 & 75.96\% & 94.50\% & 98.76\%  & 99.98\%  \\
              & 40, 40 & 93.76\% & 99.74\% & 100.00\% & 100.00\% \\
Yu-Fisher     & 5, 5   & 17.28\% & 29.30\% & 37.10\%  & 51.84\%  \\
              & 10, 10 & 41.40\% & 66.98\% & 82.16\%  & 94.56\%  \\
              & 20, 20 & 75.06\% & 94.18\% & 98.58\%  & 100.00\% \\
              & 40, 40 & 93.34\% & 99.68\% & 100.00\% & 100.00\% \\
gPCA          & 5, 5   & 18.24\% & 32.88\% & 43.44\%  & 59.98\%  \\
              & 10, 10 & 33.54\% & 66.74\% & 85.44\%  & 96.24\%  \\
              & 20, 20 & 58.00\% & 92.08\% & 98.90\%  & 100.00\% \\
              & 40, 40 & 76.24\% & 99.10\% & 99.98\%  & 100.00\% \\
HN-\texttt{mean\_test} & 5, 5   & 22.94\% & 37.26\% & 47.34\%  & 62.94\%  \\
              & 10, 10 & 49.70\% & 75.32\% & 88.40\%  & 96.92\%  \\
              & 20, 20 & 80.78\% & 96.38\% & 99.30\%  & 100.00\% \\
              & 40, 40 & 95.52\% & 99.88\% & 100.00\% & 100.00\% \\
HN-\texttt{cov\_test}  & 5, 5   & 7.66\%  & 8.34\%  & 7.56\%   & 7.82\%   \\
              & 10, 10 & 7.38\%  & 6.88\%  & 7.38\%   & 7.08\%   \\
              & 20, 20 & 6.44\%  & 6.50\%  & 6.06\%   & 6.36\%   \\
              & 40, 40 & 5.68\%  & 5.46\%  & 6.04\%   & 5.36\%   \\
Yu-\texttt{mean\_test} & 5, 5   & 23.36\% & 37.72\% & 47.82\%  & 63.24\%  \\
              & 10, 10 & 49.76\% & 75.42\% & 88.46\%  & 97.00\%  \\
              & 20, 20 & 80.80\% & 96.38\% & 99.30\%  & 100.00\% \\
              & 40, 40 & 95.52\% & 99.88\% & 100.00\% & 100.00\% \\
Yu-\texttt{cov\_test}  & 5, 5   & 2.84\%  & 3.10\%  & 2.60\%   & 2.60\%   \\
              & 10, 10 & 4.34\%  & 4.36\%  & 4.06\%   & 3.86\%   \\
              & 20, 20 & 5.08\%  & 5.10\%  & 4.60\%   & 4.94\%   \\
              & 40, 40 & 4.92\%  & 4.70\%  & 5.10\%   & 4.88\%
                \\
    \bottomrule
    \end{tabular}
    \label{Table 6}
\end{table}

\begin{table}[H]
    \centering
    \caption{Empirical powers for the alternative hypothesis $H_\mathrm{c}$}
    \begin{tabular}{cccccc}
    \toprule
    \multirow{2}{*}{Method} & \multirow{2}{*}{$n_1,n_2$} & \multicolumn{4}{c}{$p$} \\
                            \cmidrule{3-6}
                            &                        & 50     & 100    & 250    & 500    \\
    \midrule
HN            & 5, 5   & 21.88\%  & 24.16\%  & 25.88\%  & 26.40\%  \\
              & 10, 10 & 78.48\%  & 82.32\%  & 84.10\%  & 84.68\%  \\
              & 20, 20 & 99.92\%  & 100.00\% & 100.00\% & 100.00\% \\
              & 40, 40 & 100.00\% & 100.00\% & 100.00\% & 100.00\% \\
Yu-Cauchy     & 5, 5   & 25.38\%  & 27.10\%  & 25.46\%  & 25.04\%  \\
              & 10, 10 & 89.88\%  & 94.28\%  & 97.22\%  & 97.62\%  \\
              & 20, 20 & 99.94\%  & 100.00\% & 100.00\% & 100.00\% \\
              & 40, 40 & 100.00\% & 100.00\% & 100.00\% & 100.00\% \\
Yu-Fisher     & 5, 5   & 20.12\%  & 20.88\%  & 22.34\%  & 21.64\%  \\
              & 10, 10 & 87.36\%  & 92.40\%  & 95.66\%  & 96.92\%  \\
              & 20, 20 & 99.96\%  & 100.00\% & 100.00\% & 100.00\% \\
              & 40, 40 & 100.00\% & 100.00\% & 100.00\% & 100.00\% \\
gPCA          & 5, 5   & 6.46\%   & 7.20\%   & 6.84\%   & 6.86\%   \\
              & 10, 10 & 6.58\%   & 5.58\%   & 6.04\%   & 5.64\%   \\
              & 20, 20 & 5.58\%   & 5.02\%   & 4.82\%   & 5.44\%   \\
              & 40, 40 & 5.34\%   & 5.40\%   & 5.02\%   & 5.06\%   \\
HN-\texttt{mean\_test} & 5, 5   & 9.40\%   & 9.20\%   & 8.20\%   & 7.92\%   \\
              & 10, 10 & 8.22\%   & 6.80\%   & 6.86\%   & 6.46\%   \\
              & 20, 20 & 7.06\%   & 6.14\%   & 5.84\%   & 6.38\%   \\
              & 40, 40 & 7.02\%   & 6.98\%   & 6.02\%   & 5.90\%   \\
HN-\texttt{cov\_test}  & 5, 5   & 45.94\%  & 50.42\%  & 51.86\%  & 52.48\%  \\
              & 10, 10 & 93.82\%  & 97.16\%  & 98.92\%  & 99.20\%  \\
              & 20, 20 & 99.96\%  & 100.00\% & 100.00\% & 100.00\% \\
              & 40, 40 & 100.00\% & 100.00\% & 100.00\% & 100.00\% \\
Yu-\texttt{mean\_test} & 5, 5   & 9.52\%   & 9.36\%   & 8.34\%   & 7.98\%   \\
              & 10, 10 & 8.22\%   & 6.84\%   & 6.88\%   & 6.52\%   \\
              & 20, 20 & 7.08\%   & 6.14\%   & 5.84\%   & 6.38\%   \\
              & 40, 40 & 7.02\%   & 6.96\%   & 6.04\%   & 5.90\%   \\
Yu-\texttt{cov\_test}  & 5, 5   & 38.58\%  & 41.82\%  & 43.18\%  & 43.90\%  \\
              & 10, 10 & 92.84\%  & 96.56\%  & 98.68\%  & 99.04\%  \\
              & 20, 20 & 99.96\%  & 100.00\% & 100.00\% & 100.00\% \\
              & 40, 40 & 100.00\% & 100.00\% & 100.00\% & 100.00\%
                \\
    \bottomrule
    \end{tabular}
    \label{Table 7}
\end{table}

\begin{table}[H]
    \centering
    \caption{Empirical powers for the alternative hypothesis $H_\mathrm{m} \cap H_\mathrm{c}$}
    \begin{tabular}{cccccc}
    \toprule
    \multirow{2}{*}{Method} & \multirow{2}{*}{$n_1,n_2$} & \multicolumn{4}{c}{$p$} \\
                            \cmidrule{3-6}
                            &                        & 50     & 100    & 250    & 500    \\
    \midrule
HN-\texttt{mean\_test} & 5, 5   & 16.62\%  & 21.94\%  & 26.10\%  & 32.24\%  \\
              & 10, 10 & 25.88\%  & 39.48\%  & 52.62\%  & 68.28\%  \\
              & 20, 20 & 54.22\%  & 80.26\%  & 92.44\%  & 98.54\%  \\
              & 40, 40 & 82.50\%  & 97.78\%  & 99.84\%  & 100.00\% \\
HN-\texttt{cov\_test}  & 5, 5   & 46.24\%  & 50.52\%  & 51.76\%  & 52.46\%  \\
              & 10, 10 & 93.88\%  & 97.08\%  & 98.96\%  & 99.20\%  \\
              & 20, 20 & 99.96\%  & 100.00\% & 100.00\% & 100.00\% \\
              & 40, 40 & 100.00\% & 100.00\% & 100.00\% & 100.00\% \\
Yu-\texttt{mean\_test} & 5, 5   & 16.76\%  & 21.96\%  & 26.34\%  & 32.82\%  \\
              & 10, 10 & 26.00\%  & 39.60\%  & 52.78\%  & 68.42\%  \\
              & 20, 20 & 54.30\%  & 80.26\%  & 92.44\%  & 98.54\%  \\
              & 40, 40 & 82.52\%  & 97.78\%  & 99.84\%  & 100.00\% \\
Yu-\texttt{cov\_test}  & 5, 5   & 38.50\%  & 42.34\%  & 43.06\%  & 43.70\%  \\
              & 10, 10 & 92.96\%  & 96.56\%  & 98.68\%  & 99.04\%  \\
              & 20, 20 & 99.96\%  & 100.00\% & 100.00\% & 100.00\% \\
              & 40, 40 & 100.00\% & 100.00\% & 100.00\% & 100.00\% \\
    \bottomrule
    \end{tabular}
    \label{Table 8}
\end{table}

\begin{table}[H]
    \centering
    \caption{Median computational time (in milliseconds)}
    \begin{tabular}{cccccc}
    \toprule
    \multirow{2}{*}{Method} & \multirow{2}{*}{$n_1,n_2$} & \multicolumn{4}{c}{$p$} \\
                            \cmidrule{3-6}
                            &                        & 50     & 100    & 250    & 500    \\
    \midrule
HN        & 5, 5   & 1.09    & 5.38    & 61.51   & 500.38  \\
          & 10, 10 & 1.11    & 5.45    & 62.47   & 503.75  \\
          & 20, 20 & 1.41    & 6.24    & 65.83   & 517.17  \\
          & 40, 40 & 2.12    & 7.66    & 74.54   & 545.02  \\
Yu-Cauchy & 5, 5   & 4.67    & 8.23    & 19.41   & 37.92   \\
          & 10, 10 & 4.71    & 8.42    & 19.83   & 39.68   \\
          & 20, 20 & 4.94    & 9.20    & 22.16   & 44.38   \\
          & 40, 40 & 6.42    & 11.84   & 28.13   & 55.36   \\
Yu-Fisher & 5, 5   & 4.66    & 8.28    & 19.76   & 42.31   \\
          & 10, 10 & 4.70    & 8.57    & 20.69   & 42.40   \\
          & 20, 20 & 5.06    & 9.31    & 23.58   & 45.77   \\
          & 40, 40 & 6.36    & 11.87   & 28.13   & 55.27   \\
gPCA      & 5, 5   & 703.78  & 703.25  & 729.38  & 753.38  \\
          & 10, 10 & 752.96  & 752.59  & 807.17  & 842.65  \\
          & 20, 20 & 876.19  & 889.93  & 959.71  & 1034.76 \\
          & 40, 40 & 1149.69 & 1166.73 & 1284.06 & 1356.28 \\
    \bottomrule
    \end{tabular}
    \footnotesize
    
    Randomly generate two groups of samples by the scheme for the alternative hypothesis $H_\mathrm{m} \cap H_\mathrm{c}$, and repeat each $(n_1,n_2,p)$ setting 99 times.
    \label{Table 9}
\end{table}

\section*{Acknowledgments}
The author would like to sincerely thank Huiru Tang's laboratory (Human Phenome Institute, Fudan University) for providing the server resources and expert guidance in metabolomics.

\end{document}